\documentclass[aps,prl,twocolumn,a4paper,superscriptaddress,10pt]{revtex4-2}
\usepackage{etoolbox}
\apptocmd{\sloppy}{\hbadness 2000\relax}{}{}

\usepackage[english]{babel}
\usepackage{graphicx}
\usepackage{hyperref}
\hypersetup{
    colorlinks,
    linkcolor={red!50!black},
    citecolor={blue!50!black},
    urlcolor={blue!80!black}}
\usepackage{cancel}
\usepackage{amsmath}
\usepackage{amsfonts}
\usepackage{amssymb}
\usepackage{bm}
\usepackage[normalem]{ulem}
\usepackage{physics}
\usepackage{relsize}
\usepackage{mathtools}
\usepackage{indentfirst}
\usepackage{pstricks}
\usepackage{tikz}
\tikzset{>=latex}
\usetikzlibrary{patterns}
\usepackage[overload]{empheq}
\usepackage{xcolor}
\usepackage{natbib}
\usepackage{dsfont}
\usepackage{microtype}
\usepackage{mathrsfs}

\usepackage[caption=false]{subfig}

\usepackage{color}

\makeatletter
\renewcommand{\fnum@figure}{FIG. \thefigure}
\makeatother

\definecolor{MatBlue}{rgb}{0.400822, 0.522007, 0.85}
\definecolor{MatOrange}{rgb}{0.922526, 0.385626, 0.209179}

\begin{document}
\title{Stress Isotropization in Weakly Jammed Granular Packings}
\author{Félix Benoist}
%\email{felix.benoist@gimm.pt}
\affiliation{Université Paris-Saclay, CNRS, LPTMS, 91400, Orsay, France}
\affiliation{Gulbenkian Institute of Molecular Medicine, Oeiras, Portugal}
\author{Mehdi Bouzid}
\affiliation{Université Grenoble Alpes, CNRS, Grenoble INP, 3SR, 38000 Grenoble, France}
\author{Martin Lenz}
\email{martin.lenz@cnrs.fr}
\affiliation{Université Paris-Saclay, CNRS, LPTMS, 91400, Orsay, France}
\affiliation{PMMH, CNRS, ESPCI Paris, PSL University, Sorbonne Université,
Université de Paris, F-75005, Paris, France}

\begin{abstract}
When sheared, granular media experience localized plastic events known as shear transformations, which generate anisotropic internal stresses. Under strong confining pressure, the response of granular media to local force multipoles is essentially linear, resulting in quadrupolar propagated stresses. This can lead to additional plastic events along the direction of increase in relative stress. Closer to the unjamming transition, however, as the confining pressure and the shear modulus vanish, nonlinearities become relevant. Yet, the consequences of these nonlinearities on the stress response to plastic events remain poorly understood. We show with granular dynamics simulations that this brings about an isotropization of the propagated stresses, in agreement with a previously developed continuum elastic model. This could significantly modify the yielding transition of weakly jammed amorphous media, which has been conceptualized as an avalanche of such plastic events.
\end{abstract}

\maketitle

{\bf Introduction.}
% Yielding stems from microscopic events
When subjected to large enough macroscopic shear stress, amorphous solids such as granular packings, foams, metallic glasses or toothpaste start to flow~\cite{Nicolas18,Bonn17}. This yielding transition originates in microscopic events where the material locally undergoes a plastic deformation~\cite{Argon79}. Each of these so-called shear transformations induces new microscopic stresses in its surroundings, which can then trigger further shear transformations. Above a critical macroscopic stress, the catastrophic accumulation of such events is widely believed to cause the whole material to yield and transition from a solid-like to a fluid-like behavior~\cite{Divoux24}.

\begin{figure}[!b]
    \centering
    \includegraphics[width=\linewidth]{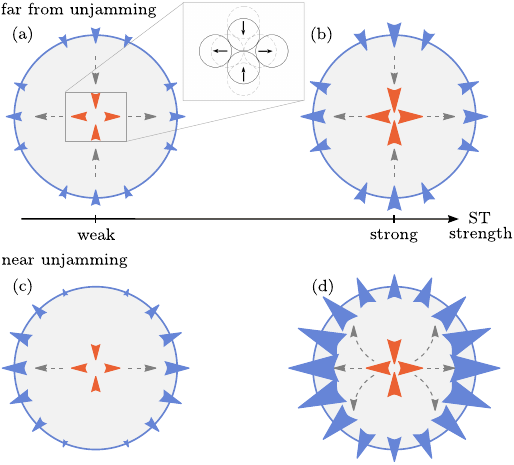}
   \caption{
   {\bf The far-field stresses induced by a shear transformation (ST) become more isotropic near unjamming.}
   (a)~In an amorphous medium, a shear transformation, \textit{e.g.}, a local change of neighbors between grains (inset)~\cite{Kabla03,Desmond15}, applies a local force dipole (orange arrowheads) on the surrounding medium. These forces propagate through the medium (grey arrows), resulting in stresses at the medium's boundary (blue arrowheads).
   (b)~Far from unjamming, the medium propagates stresses according to linear elasticity. The symmetry of this stress response is thus independent of the magnitude of the local forces. 
   (c)~Close to unjamming, the medium may not support the propagation of tensile stresses, resulting in a dilational stress response.
   (d)~For large local forces, stress redistribution within the medium results in an increasingly isotropic dilation.}
\label{fig:4sketches}
\end{figure}

% Characteristics of yielding derive from propagator
This yielding transition has been widely studied via mesoscopic elasto-plastic models~\cite{Hebraud98,Bocquet09,Nicolas18,Lin_PNAS14,Budrikis17, Talamali11,Bouzid15}. Its universality class depends on the propagator, which determines how a shear transformation redistributes stress. Linear elastic solids display a so-called Eshelby propagator, whose quadrupolar symmetry imposes a balance of positive and negative stresses~\cite{Eshelby57,Karimi18}. This symmetry, associated with the dipolar deformation field illustrated in Fig.~\ref{fig:4sketches}(a,b), dictates the critical exponents at the yielding transition. Among them, the Herschel-Bulkley exponent describes how abruptly the material starts to flow once the macroscopic stress exceeds its critical value~\cite{Lin_PNAS14,Lin_EPL14}. Phenomena such as shear-banding and aging have also been linked to the form of the Eshelby propagator~\cite{Dasgupta12,Tyukodi16_yield,Martens12}.

% Eshelby kernel may not apply close to unjamming
Although the validity of the Eshelby propagator is well established in dense amorphous solids~\cite{Tanguy06}, looser packings may display more complex responses. Experiments on weakly jammed emulsions~\cite{Liu98,vanHecke09} thus indicate a non-Eshelby propagation, whereby the change in local stress surrounding a shear transformation has the same sign in all directions~\cite{Desmond15}. Similarly, an isotropic core is observed in the displacement response to force dipoles of simulated harmonic sphere packings near unjamming~\cite{Giannini24}. In this article, we propose that such deviations from the Eshelby propagator are generically expected for amorphous solids close to unjamming.

% Local asymmetric material response may modify kernel and ultimately macroscopic properties (including paper outline)
Our approach is based on the observation that as an isotropic material approaches unjamming, one or both elastic moduli vanish, while its higher-order nonlinear elastic response remains finite~\cite{O'Hern03, DagoisBohy17,vanDeen14}, leading to a nonlinear response to local shear transformations. To understand the origin of this nonlinearity, consider two contacting grains within the medium. Compressive forces maintain contact, whereas tensile forces tend to pull the grains apart, favoring the transmission of compressive over tensile stresses. As a result, local forcing induces a bias toward isotropic dilation [Fig.~\ref{fig:4sketches}(c,d)], an effect that we have previously termed rectification~\cite{ronceray_fiber_2016,Benoist23}. Here, we validate this isotropization using numerical simulations of 2D granular packings. We find that the system's tendency to isotropization diverges as unjamming is approached, in agreement with a nonlinear elastic model~\cite{Benoist23}. Finally, a simple toy model suggests that this effect can alter critical exponents and qualitative features of the yielding transition in amorphous solids near unjamming.
\newline

\begin{figure}[!t]
    \centering
    \includegraphics[width=\linewidth]{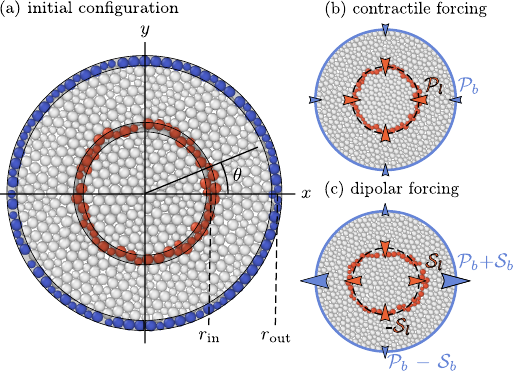}
   \caption{{\bf We subject circular jammed packings to small internal forces.}
   (a)~Packing of the type used in our simulations but with fewer disks. We exert radial forces on the disks in the shaded region near $r_\text{in}$ (orange), and measure the forces exerted on the disks in the shaded region near $r_\text{out}$ (blue).
   (b)~The same packing under isotropic contractile forcing, $\mathcal P_l<0$, corresponding to a local shrinkage of the original orange ring. In the final configuration, some gray disks are now subject to the forcing, and some orange ones are not. The dashed circle has radius $r_\text{in}$. (c)~Dipolar forcing, $\mathcal S_l>0$.}
\label{fig:force_ring}
\end{figure}

{\bf Stress propagation around a shear transformation near unjamming.}
We consider packings of frictionless, bidisperse disks confined within a circular arena of radius $r_\text{out}$, at mechanical equilibrium, as described in the End Matter. The two disk species are present in equal proportions, with a diameter ratio of 1.4. We use the mean diameter as our length unit. The disk area fraction $\phi$ is set slightly above the critical value $\phi_c\simeq0.84$ at which the packing unjams~\cite{O'Hern03}, such that $\Delta\phi=\phi-\phi_c\lesssim0.2$. A pair of disks with overlap $\delta$ interacts elastically via a Hertzian potential proportional to $\delta^{5/2}$~\cite{Makse04}. The associated stiffness sets our unit of stress. For the values of $\phi$ considered here, the resulting initial pressure is low, with $P_\text{init}\lesssim 0.02$. To mimic stress propagation around a shear transformation, we apply internal forces at a radius $r_\text{in}$ [Fig.~\ref{fig:force_ring}(a)]. In practice, we use mesoscopic values for $r_\text{in}$ to mitigate fluctuations due to the medium's disorder, and we apply forces of the order of, or smaller than, $P_\text{init}$ to avoid triggering extensive plastic reorganizations~\cite{Bouzid15}. 

We characterize the magnitude and anisotropy of these forces through the corresponding coarse-grained local stress tensor $\bar{\bm\sigma}^l$ [definition in End Matter]. We monitor stress propagation via the boundary stress response tensor $\bar{\bm\sigma}^b$, which characterizes the forces exerted by the medium at its boundary in response to $\bar{\bm\sigma}^l$. Our local forcing is decomposed into an isotropic pressure $\mathcal P_l$ and a shear stress $\mathcal S_l$, respectively illustrated in Figs.~\ref{fig:force_ring}(b) and (c). We decompose $\bar{\bm\sigma}^b$ similarly. Placing ourselves in the eigenbasis of tensor $\bar{\bm\sigma}^l$, we write
\begin{equation}\label{eq:bar_sig}
	\bar{\bm\sigma}^i = -\mathsmaller{\begin{pmatrix}\mathcal P_i+\mathcal S_i&0\\0&\mathcal P_i-\mathcal S_i\end{pmatrix}}
\end{equation}
for $i\in\lbrace l,b\rbrace$. Qualitatively, a positive $\mathcal P_b$ corresponds to an overall dilation of the medium. Note that for an individual realization of our granular packing, the off-diagonal components of $\bar{\bm\sigma}^b$ in Eq.~\eqref{eq:bar_sig} may not vanish. Symmetry however imposes that their average over the disorder does, and in practice the off-diagonal components do not exceed $10\%$ of the diagonal ones for any of our individual packings.

\begin{figure}[!t]
    \centering
    \includegraphics[width=\linewidth]{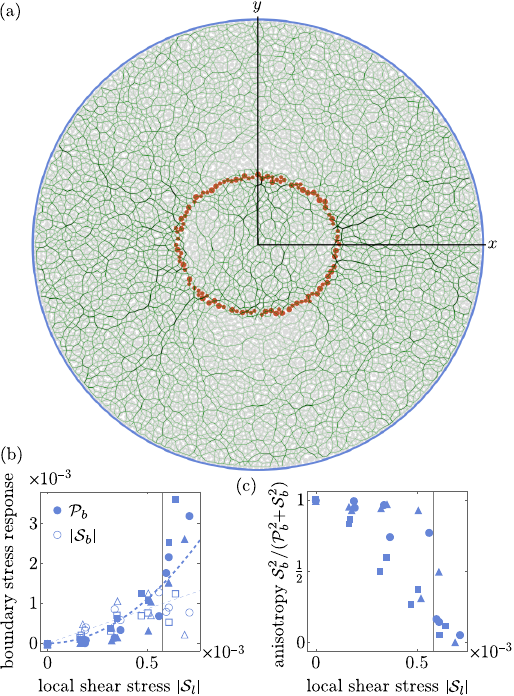}
   \caption{{\bf Force chains rearrange to create boundary dilation out of local shear stress.} (a)~Local shear stress rearranges force chains in a packing. The green segments have widths and colors proportional to the forces between neighboring disks. We refer to consecutive segments with large widths as force chains. Here, we have $\simeq 6700$ disks, $r_\text{out}\simeq 44$, $r_\text{out}/r_\text{in}=3$, $\Delta\phi\simeq0.03$. 
   (b)~Dilation under dipolar forcing for three initial configurations (circles, squares and triangles), demonstrating reproducibility. We fit the data using $\mathcal P_b=\alpha\mathcal S_l^2$ with $\alpha\simeq 4400$, and $\mathcal S_b=(1+B)\mathcal S_l$ with $B\simeq0.7$. Here, $r_\text{out}/r_\text{in}=8$, $r_\text{in}\simeq 5.5$, $\Delta\phi\simeq 0.03$.
   (c)~The anisotropy in the stress response decreases as the local shear stress increases. The vertical line indicates the local shear stress at which this anisotropy falls under 1/2.} 
\label{fig:gran_dip}
\end{figure}

In the initial configurations, the confining pressure elicits an isotropic arrangement of force chains, see Fig.~S1 in~\cite{SM}. The application of a dipolar forcing $\mathcal S_l>0$ rearranges the force chains anisotropically as shown in Figs.~\ref{fig:gran_dip}(a) and~S2-S3. We focus on the outer region $r>r_\text{in}+\tfrac{1}{2}$ through which the local forcing propagates to the boundary. As $\mathcal S_l$ increases and nonlinearities become prevalent, the force chains in the vicinity of the $y$ axis near radius $r_\text{in}$ become weaker. Conversely, the force chains near the $x$ axis are reinforced, both horizontally and along directions with significant angles with the $x$ axis. At large values of $r$, this anisotropic propagation leads to reinforced force chains even close to the $y$ axis. As a result, a dilational boundary pressure $\mathcal P_b>0$ emerges in response to the local shear stress $\mathcal S_l$ [Figs.~\ref{fig:gran_dip}(b) and~S4]. In extreme cases, this can rectify the stress response toward dilation in all directions, thus making it more isotropic; see Fig.~\ref{fig:gran_dip}(c). Under isotropic forcing $\mathcal P_l\neq0$, we also observe nonlinear relationships between $\mathcal P_b$ and $\mathcal P_l$ [Figs.~S5-S6]. Overall, this minimal new setup clearly shows how microscopic force chain rearrangements induce the isotropization of propagated stresses illustrated in Fig.~\ref{fig:4sketches}.
\newline

{\bf Dependence of the non-Eshelby propagation on material behavior.}
To assess the relationship between isotropization and unjamming, we now turn to theory. Within the linear regime, homogeneous elastic media propagate stress without alteration, such that $\mathcal P_b=\mathcal P_l$ and $\mathcal S_b=\mathcal S_l$. For weak nonlinearities, symmetries dictate
\begin{equation}\label{eq:PB-SB}
    \mathcal P_b\sim \mathcal{P}_l + \alpha\mathcal S_l^2 + \beta \mathcal{P}_l^2 \qq{and} \mathcal S_b\propto\mathcal S_l.
\end{equation}
The term $\alpha\mathcal S_l^2$, if large and positive, induces dilational stresses that come to dominate the medium's response and is thus responsible for isotropization. In a previous paper~\cite{Benoist23}, we expressed $\alpha$ as a function of the material's elastic properties. To parameterize those, we consider a jammed disk packing initially at 
area fraction $\phi_0=\phi_c+\Delta\phi$ with differential bulk and shear moduli $\kappa_0$ and $\mu_0$. When subjected to a small additional compression $\delta\phi\ll \Delta\phi$, its moduli become, to first order in $\delta\phi$,
\begin{equation}\label{eq:K,G_wnl}
	\kappa= \kappa_0\left(1-\kappa_1\frac{\delta\phi}{\phi_c}\right) \qq{and} \mu= \mu_0\left(1-\mu_1\frac{\delta\phi}{\phi_c}\right).
\end{equation}
The isotropization coefficient $\alpha$ then reads
\begin{equation}\label{eq:alpha}
	\alpha=-\frac{1}{\mu_0}\left[\left(\kappa_1+\frac{3}{2}\right)\alpha_1+\left(\mu_1+\frac{3}{2}\right)\alpha_2\right],
\end{equation}
where $\alpha_1$ and $\alpha_2$ are positive functions of the ratio of radii $r_\text{out}/r_\text{in}$ and Poisson's ratio $\nu=(\kappa_0-\mu_0)/(\kappa_0+\mu_0)$; see End Matter for full expressions.

For granular materials under Hertzian interactions, the bulk and shear moduli vanish at unjamming as $\kappa\sim \kappa_0\approx\sqrt{\Delta\phi}$ and $\mu\sim\mu_0\approx\Delta\phi$~\cite{O'Hern03}. Expanding the expressions of $\kappa$ and $\mu$ for small $\delta\phi$ and equating the result to Eq.~\eqref{eq:K,G_wnl} yields
\begin{equation}\label{eq:k1,mu1}
	\kappa_1 \underset{\Delta\phi\rightarrow 0}{\sim} -\frac{1}{2}\frac{\phi_c}{\Delta\phi} \qq{and} \mu_1 \underset{\Delta\phi\rightarrow 0}{\sim} -\frac{\phi_c}{\Delta\phi}.
\end{equation}
Therefore, $\kappa_1$ and $\mu_1$ are both negative. Granular media indeed soften under tension ($\delta\phi<0$) and stiffen under compression. These nonlinear coefficients moreover diverge near unjamming as $(\Delta\phi)^{-1}$, leading to a large positive isotropization coefficient $\alpha$, see End Matter. According to Eq.~\eqref{eq:PB-SB}, an anisotropic forcing already gives rise to a significantly isotropized far-field stress response (\emph{i.e.}, $\mathcal{P}_b > \mathcal{S}_b$) for a small $\mathcal{S}_l \approx \alpha^{-1}\approx(\Delta\phi)^2$.

\begin{figure}[!t]
    \centering
    \includegraphics[width=\linewidth]{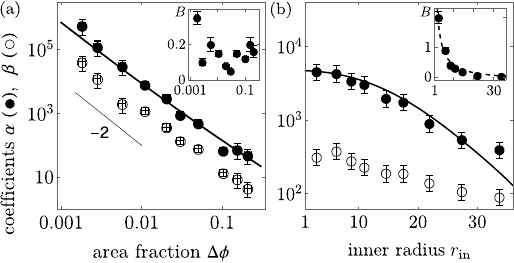}
   \caption{{\bf Isotropization prevails close to unjamming and in large systems.}
   (a)~Log-log plot of coefficients $\alpha$ and $\beta$ obtained as in the fits of Fig.~\ref{fig:gran_dip}(b) showing a $(\Delta\phi)^{-2}$ divergence as unjamming is approached. The black line shows our theoretical predictions [Eq.~\eqref{eq:alpha}] and we observe $\beta\simeq\alpha/9$. \emph{Inset:} the phenomenological coefficient $B$ does not strongly depend on $\Delta\phi$.
   Here, $r_\text{in}\simeq22$ and $r_\text{out}\simeq44$.
   (b)~Holding the outer radius $r_\text{out}\simeq44$ constant, the isotropization coefficient $\alpha$ is strongest for small $r_\text{in}$, in agreement with the theoretical prediction (End Matter). \emph{Inset:} $B$ is also large for small $r_\text{in}$; the dashed line shows a heuristic dependence $B=(r_\text{out}/r_\text{in}-1)/8$.
   Here, $\Delta\phi\simeq0.03$. Bars show standard deviation across three simulations.}
\label{fig:gran_biases}
\end{figure}

To confirm this predominance of isotropization in the vicinity of unjamming, we measure coefficients $\alpha$ and $\beta$ in simulations of the type of those presented in Fig.~\ref{fig:gran_dip} for different values of $\Delta\phi$ and $r_\text{in}$ (additional fits in Figs.~S4-S9) and report them in Fig.~\ref{fig:gran_biases}. We find that the small-nonlinearity expansion of Eq.~\eqref{eq:PB-SB} describes our data well even in regimes where the nonlinear terms are comparable to or larger than the linear ones. We observe an unexpected steeper linear dependence $\mathcal{S}_b=(1+B)\mathcal{S}_l$ than predicted, but find that the phenomenological coefficient $B>0$ is not large enough to prevent isotropic dilation from dominating the response of our packings (insets of Fig.~\ref{fig:gran_biases}). Both coefficients $\alpha$ and $\beta$ are positive, and thus contribute to the medium's dilation in the nonlinear regime. This dilation is moreover strongest for small $\Delta\phi$ and for a very localized forcing ($r_\text{in}$ small). Finally, we compare the values of the main isotropization coefficient $\alpha$ with the prediction of Eqs.~(\ref{eq:alpha}-\ref{eq:k1,mu1}). We find a very good agreement without any adjustable parameters (we obtain $\kappa_0$ and $\mu_0$ from Ref.~\cite{O'Hern03}), confirming that close to unjamming, the weakening of the packing's linear response induces an overwhelmingly dilational, non-Eshelby response to localized forces. 
\newline

{\bf Macroscopic implications of the non-Eshelby propagation.} 
In amorphous solids, shear transformations can trigger further plastic events via stress propagation. In elasto-plastic models, this cascade is governed by the Eshelby propagator’s quadrupolar symmetry, which makes a shear transformation equally likely to promote or suppress plastic reorganizations in nearby regions. Loosely defining a scalar stress $\sigma$ characterizing how far a region is from the reorganization threshold $\sigma_c$, this implies that $\sigma$ undergoes a symmetric random walk prior to reaching $\sigma_c$. By contrast, Eq.~\eqref{eq:PB-SB} predicts an additional dilational stress $\mathcal{P}_b$, which should bias this walk and alter the universality of the fluidization transition.

To illustrate the macroscopic implications of this breaking of the stress-reversal symmetry, we turn to a simple mean-field elasto-plastic model~\cite{Hebraud98} that monitors the probability density $P(\sigma, t)$ of local stresses over time $t$. In a material under external shear at rate $\dot{\gamma}$, $\sigma$ increases at rate $\mu\dot{\gamma}$. It also diffuses due to stress kicks from distant shear transformations, with diffusion coefficient $D=a\Gamma$, where $\Gamma$ is the number of plastic reorganizations per unit time and $a$ a constant. We reason that the added dilational stress $\mathcal{P}_b$ increases the packing pressure and thus reduces the tendency to yield~\cite{Bonn17}. At leading order, this introduces a drift away from the yielding threshold that is proportional to $\Gamma$. It can thus be represented by adding a current $-bD$ to the evolution equation for $P(\sigma,t)$, where $b$ is a constant:
\begin{equation}\label{eq:Hebraud-Lequeux}
    \partial_tP=-(\mu\dot{\gamma}-ba\Gamma)\partial_\sigma P+a\Gamma\partial_\sigma^2P-\nu(\sigma)P+\Gamma\delta(\sigma).    
\end{equation}
Here, the disappearance rate $\nu(\sigma)=\tau^{-1}H(\abs{\sigma}-\sigma_c)$, where $H$ denotes the Heaviside step function, implies that each region whose stress exceeds the critical value $\sigma_c$ undergoes a plastic reorganization and is removed from the system. It is then reintroduced as a new stress-less configuration through the last term of Eq.~\eqref{eq:Hebraud-Lequeux} involving Dirac's delta function $\delta$, with the condition that $\Gamma(t)=\int_{-\infty}^{+\infty}\nu(\sigma)P(\sigma,t)\,\text{d}\sigma$. As $b$ quantifies the relative importance of the dilational and Eshelby stress propagation, we expect it to become relevant close to unjamming.

Analyzing Eq.~\eqref{eq:Hebraud-Lequeux} in a steady state (details in~\cite{SM}) reveals that just like the classical $b=0$ case, our extended $b\neq 0$ model displays an unjamming transition from a solid-like phase devoid of plastic events at zero shear rate ($\Gamma=0$) to a fluid-like phase ($\Gamma\neq 0$) upon an increase of $a$ through a critical value. At the transition, the rheology of the material is described by a Herschel-Bulkley exponent of 1/2:
\begin{equation}\label{eq:rheology}
    \expval{\sigma}-\expval{\sigma(\dot{\gamma}=0)}\approx\dot{\gamma}^{1/2}.
\end{equation}
By contrast, in the Eshelby-like ($b=0$) case this dependence is $\expval{\sigma}\approx\dot{\gamma}^{1/5}$ and only crosses over to a Herschel-Bulkley exponent of 1/2 deeper in the jammed phase~\cite{Agoritsas15}. While derived in a simplistic model, this indicates that the loss of the Eshelby-like symmetry can have macroscopic implications for the rheology of amorphous materials.
\newline

{\bf Discussion.}
Our study sheds light on the transmission of internally generated stresses in granular systems close to unjamming. Many elasto-plastic models assume that, following a shear transformation, this transmission is well described by an Eshelby-like linear elasticity kernel~\cite{Nicolas18,Merabia16,Lin_PNAS14}. We show that elastic nonlinearities inherent to the unjamming transition instead lead to substantial isotropic dilational stresses around local rearrangements. This isotropization is strongest for rearrangements spanning a few particle diameters, comparable to that of shear transformations~\cite{Amon12}. This mirrors earlier experimental~\cite{Desmond15} and numerical~\cite{Giannini24} findings.

Although our analysis focuses on Hertzian disk packings, we expect similar results for harmonic interactions [End Matter] and random spring networks~\cite{Lerner14,Ellenbroek_EPL09}. This universal character is reflected in the good agreement between our simulations and a continuum theory devoid of microscopic assumptions; see~\cite{SM} for a study of finite-size effects and plasticity (and Refs.~\cite{Silbert05,Karimi15,Ellenbroek_PRE09,Wyart05,Dinkgreve15,Barrat11,Maloney08} therein). While based on a small-stress, weakly nonlinear expansion, this formalism quantitatively predicts isotropized stresses even in regimes where they are significantly larger than the stresses predicted by linear elasticity. This is reminiscent of successful predictions of the onset of failure in amorphous solids based on lowest-order nonlinearities~\cite{Karmakar10}. Crucially, this isotropization requires the vanishing of at least one elastic modulus at unjamming. Systems where $K$ and $G$ remain finite at the transition should thus display negligible far-field stress isotropization.

We use an elasto-plastic toy model to bring out the macroscopic consequences of the isotropization-induced breaking of symmetry between positive and negative stresses. Consistent with recent non-mean-field results, we find that this asymmetry changes the characteristics of the unjamming transition~\cite{Jocteur24}. Moreover, while yielding in strongly jammed systems tends to concentrate along transient slip lines~\cite{Maloney04,Maloney06,Dasgupta12}, we predict more homogeneous, ductile-like yielding in weakly jammed systems due to more isotropic propagated stresses~\cite{Richard20,Tyukodi16_yield}.
\newline

\begin{acknowledgments}
{\bf Acknowledgments.}
ML and FB thank Éric Clément and Sylvain Patinet for fruitful discussions. ML was supported by Marie Curie Integration Grant PCIG12-GA-2012-334053, “Investissements d’Avenir” LabEx PALM (ANR-10-LABX-0039-PALM), ANR-21-CE11-0004-02, ANR-22-ERCC-0004-01 and ANR-22-CE30-0024-01, as well as ERC Starting Grant 677532 and the Impulscience program of Fondation Bettencourt-Schueller. ML’s group belongs to the CNRS consortium AQV.
\end{acknowledgments}

%\bibliography{bibfile0}

%

%\newpage

%%%%%%%%%% Merge with end matter %%%%%%%%%%
\onecolumngrid
\begin{center}\textbf{\large End Matter}\end{center}
\twocolumngrid
%\renewcommand{\thefigure}{S\arabic{figure}}
%\renewcommand{\theHfigure}{S\arabic{figure}}
%\renewcommand{\bibnumfmt}[1]{[S#1]}
%\renewcommand{\citenumfont}[1]{S#1}
%%%%%%%%%% Prefix a "S" to all equations, figures, tables and reset the counter %%%%%%%%%%
%\vspace{5mm}

\setcounter{equation}{0}
\makeatletter
\renewcommand{\theequation}{A\arabic{equation}}
%, as described in the End Matter. 
% OR (see End Matter).
{\bf Appendix A: Simulation methods.}
We obtain equilibrated packings by using granular dynamics via LAMMPS~\cite{Thompson22} version stable\_3Mar2020 as follows. We initialize $\simeq6700$ disks with stiffness $k$ in a random unjammed configuration inside a circular arena ($k=1$ in the main text). Two disks of radii $r_1$ and $r_2$ with overlap $\delta$ exert frictionless repulsive forces of magnitude $k\sqrt{r_1r_2/(r_1+r_2)}\delta^{3/2}$. Likewise for the disks overlapping with the arena of radius $r_a=r_\text{out}+\tfrac D2$, where $D$ is the mean disk diameter ($D=1$ in the main text). The forces $\{\mathbf f^{\nu\to\mu}\}$ exerted on disk $\mu$ by its neighbors at positions $\{\mathbf r^\nu\}$ result in an elastic stress written $\sigma^\mu_{ij}=-\sum_\nu(r_i^\nu-r_i^\mu) f^{\mu\to\nu}_j$.
To reach a jammed configuration with packing fraction $\phi>\phi_c$, we increase the disk diameters and let the packing relax by thermal annealing~\cite{Schreck_PRE11}. We determine $\Delta\phi=\phi-\phi_c$ based on the values of the initial pressure after relaxation $P_\text{init}/k~\simeq0.27(\Delta\phi)^{3/2}$ and the excess contact number $Z_\text{init}-4\simeq3.3\sqrt{\Delta\phi}$~\cite{O'Hern03}.

To investigate the response of packings to local forcing, we define an inner ring as $r\in I_l=[r_\text{in}-\tfrac{D}{2},r_\text{in}+\tfrac{D}{2}]$ [orange disks in Fig.~\ref{fig:force_ring}(a)], with area $A_l= 2\pi r_\text{in}D$. In addition to the forces due to the initial pressure $P_\text{init}$, we subject each disk $\mu$ in the inner ring to a constant radial force written
\begin{equation}\label{eq:act_force}
	\mathbf f^\mu/k=\big(f_0+2f_2\cos 2\theta^\mu\big)\hat{\mathbf r}^\mu,\ \text{ for }\ r^\mu\in I_l,
\end{equation}
such that $f_2=0$ corresponds to an isotropic forcing [Fig.~\ref{fig:force_ring}(b)], while $f_0=0$ corresponds to a dipolar forcing [Fig.~\ref{fig:force_ring}(c)]. 
This forcing elicits a coarse-grained local stress proportional to the dipole of the added forces: 
\begin{equation}\label{eq:local_stress}
	\bar{\sigma}^l_{ij}=\frac{-2D}{A_l}\sum_{\mu\in I_l}f^\mu_i\,\hat r^\mu_j.
\end{equation}
We also define a boundary ring as $r\in I_b=[r_\text{out}-\tfrac{D}{2},r_\text{out}+\tfrac{D}{2}]$ [blue disks in Fig.~\ref{fig:force_ring}(a)], with area $A_b= 2\pi r_\text{out}D$, whose outer part sticks to the arena. The excess stresses on the disks in the boundary ring $\bm\sigma^\mu-\bm\sigma^{\mu,\text{init}}$ due to the forcing result in a boundary stress response
\begin{equation}\label{eq:bound_stress}
	\bar{\sigma}^b_{ij}=\frac{2}{A_b}\frac{r_\text{out}^2}{r_\text{in}^2}\sum_{k=x,y}\sum_{\mu\in I_b}\big(\sigma^\mu_{ik}-\sigma_{ik}^{\mu,\text{init}}\big)\,\hat r^\mu_k\,\hat r^\mu_j.
\end{equation}
Due to the ratio $r_\text{out}^2/r_\text{in}^2$ compensating for the dilution, the analytical prediction for linear elastic systems is $\bar{\bm\sigma}^b=\bar{\bm\sigma}^l$~\cite{Benoist23}. For moderate forcing, $f_0,f_2\lesssim P_\text{init}/k$, the local stress components defined in Eq.~\eqref{eq:bar_sig} increase with forcing as $\mathcal P_l\propto f_0$ and $\mathcal S_l\propto f_2$; see~\cite{SM}.
\newline

\setcounter{equation}{0}
\makeatletter
\renewcommand{\theequation}{B\arabic{equation}}
{\bf Appendix B: Continuum elastic model.} 
We previously estimated the isotropization coefficient $\alpha$ for a continuum elastic medium under internal dipolar forcing [Fig.~\ref{fig:force_ring}(c)], by expanding Hooke’s law to lowest nonlinear order~\cite{Benoist23}. In that framework, nonlinear corrections to the bulk and shear moduli, $\kappa$ and $\mu$, were characterized by the parameters $\kappa_1$ and $\mu_1$. Here, we compute $\kappa_1$ and $\mu_1$ directly in granular media and substitute them into the theoretical expression for $\alpha$ [Eq.\eqref{eq:alpha}] to generate the curves shown in Fig.\ref{fig:gran_biases}.

We quantify deformation at location $\mathbf x$ using the displacement gradient $\eta_{ij}=\partial u_i/\partial x_j$. For small strains, we expand the Cauchy stress $\bm\sigma$ to the lowest nonlinear order as $\sigma_{ij}=\mathcal K_{ijkl}\eta_{kl} + \mathcal L_{ijklmn}\eta_{kl}\eta_{mn}$. Within this framework, the elastic response of an isotropic and achiral medium to a combination of bulk deformation and simple shear, \textit{i.e.} $\bm\eta = \begin{pmatrix}\eta_{ii}/2&\eta_{xy}\\0&\eta_{ii}/2\end{pmatrix}$, is characterized by differential bulk and shear moduli $\kappa =\partial\sigma_{xx}/\partial\eta_{ii}$, $\mu =\partial\sigma_{xy}/\partial\eta_{xy}$ written
\begin{equation}\label{eq:moduli}\begin{aligned}
	\kappa =\kappa_0\left(1+\kappa_1\eta_{ii}\right)  \qq{and}
	\mu =\mu_0\left(1+\mu_1\eta_{ii}\right). 
\end{aligned}\end{equation}

In the setup of Fig.~\ref{fig:force_ring}, we consider a large packing of frictionless Hertzian disks with area fraction $\phi_c+\Delta\phi$. Near unjamming, \emph{i.e.} $0<\Delta\phi\ll1$, the moduli read 
\begin{equation}\label{eq:k,mu_Hertz}
    \kappa= K(\Delta\phi)^s \qq{and} \mu= M(\Delta\phi)^t,
\end{equation}
where $K\simeq0.3$ and $M\simeq0.2$ in units of the disk stiffness, $s\simeq0.5$ and $t\simeq1.0$~\cite{O'Hern03}. The area fraction $\phi_c+\Delta\phi$ corresponds to that of a system initially at the rigidity threshold $\phi_c$ subjected to a bulk compression $\eta_{ii}=-\eta_0=-\Delta\phi/\phi_c$. We then add an even smaller perturbation: $\eta_{ii}=-\eta_0-\delta\eta$, where $|\delta\eta|\ll\eta_0$ and $\delta\eta = \delta\phi/\phi_c$. This results in Eq.~\eqref{eq:K,G_wnl}:
\begin{equation}\label{eq:K,G_wnl1}\begin{aligned}
	\kappa=\kappa_0(1-\kappa_1\delta\eta)+\order{\delta\eta^2},\\
	\mu=\mu_0(1-\mu_1\delta\eta)+\order{\delta\eta^2},
\end{aligned}\end{equation}
where, at lowest order in $\Delta\phi$, the elastic parameters read 
\begin{equation}\label{eq:k0,mu0,k1,mu1}\begin{aligned}
    \kappa_0=K(\Delta\phi)^s,\qquad \kappa_1 =-s\,\phi_c/\Delta\phi, \\
    \mu_0= M(\Delta\phi)^t,\qquad \mu_1 =-t\,\phi_c/\Delta\phi,
\end{aligned}\end{equation}
as in Eq.~\eqref{eq:k1,mu1}.
Therefore, given $\phi_c\simeq0.84$, around \textit{e.g.} $\Delta\phi= 0.1,\,0.01$ or 0.001, we find respectively $\kappa_1\simeq-4,\,-40$ or $-400$, and $\mu_1\simeq-8,\,-80$ or $-800$. Poisson's ratio then reads
\begin{equation}\label{eq:nu_Hertz}
	\nu=\frac{\kappa_0-\mu_0}{\kappa_0+\mu_0}=1-2\frac{M}{K}(\Delta\phi)^{t-s}.
\end{equation}
$1-\nu$ thus scales approximately as $(\Delta\phi)^{0.5}$, such that media far from unjamming are more compressible.

\begin{figure}[!b]
    \centering
    \includegraphics[width=\linewidth]{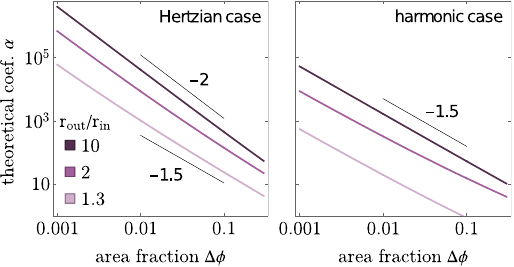}
   \caption{{\bf Theoretical prediction for the isotropization coefficient.} Equation~\eqref{eq:alpha} predicts different scalings of $\alpha$ with $\Delta\phi$ for varying values of $r_\text{out}/r_\text{in}$. For area fractions $\Delta\phi\in[10^{-3},10^{-1}]$, the exponent varies from $-2$ when $r_\text{out}/r_\text{in}=10$, to $-1.5$ when $r_\text{out}/r_\text{in}=1.3$.}
\label{bias_th}
\end{figure}

Now that the elastic parameters are properly defined, we enter them into the expression of $\alpha$ from Ref.~\cite{Benoist23} reproduced in Eq.~\eqref{eq:alpha}. Therein, $\alpha_1$ and $\alpha_2$ are positive functions of $\rho =(r_\text{out}/r_\text{in})^2$ and $\nu$ written
\begin{align*}
    \frac{X\alpha_1}{1-\nu^2}&=405-108\nu-54\nu^2+12\nu^3+\nu^4 \\
&\quad +(324-180\nu-24\nu^2-36\nu^3-4\nu^4)\rho \\
&\quad +(378-288\nu+120\nu^2+24\nu^3+6\nu^4)\rho^2 \\
&\quad +(108-180\nu+48\nu^2+12\nu^3-4\nu^4)\rho^3\\
&\quad +(81-108\nu+54\nu^2-12\nu^3+\nu^4)\rho^4
\end{align*}
and
\begin{align*}
    X\alpha_2 &=81-54\nu+351\nu^2-84\nu^3-49\nu^4+10\nu^5+\nu^6 \\
&\quad -(684\nu-204\nu^2+120\nu^3 +8\nu^4+28\nu^5+4\nu^6)\rho \\
&\quad +(594-900\nu+1122\nu^2-360\nu^3 \\
&\quad +102\nu^4+12\nu^5+6\nu^6)\rho^2 \\
&\quad +(216-1116\nu+924\nu^2-312\nu^3 \\
&\quad +16\nu^4+20\nu^5-4\nu^6)\rho^3 \\
&\quad + (405-702\nu+567\nu^2-276\nu^3 \\
&\quad +83\nu^4-14\nu^5+\nu^6)\rho^4,
\end{align*}
where
\begin{align*}
    X&=4\frac{\rho}{\rho-1}(3-\nu)^2\big[2(3+\nu)+(3-\nu)(\rho+\rho^2)\big]^2.
\end{align*}
Given the dependencies of the elastic parameters with the area fraction $\Delta\phi$ in Eq.~\eqref{eq:k0,mu0,k1,mu1}, approximating $s$ to 0.5 and $t$ to 1, we can expand Eq.~\eqref{eq:alpha} for small $\Delta\phi$ as
\begin{equation}
	\alpha\sim\alpha^{(0)}(\Delta\phi)^{-2}+\alpha^{(1)}(\Delta\phi)^{-3/2}, %+\order{(\Delta\phi)^{-1}},
\end{equation}
where 
\begin{align}
	\hspace{-1mm}\alpha^{(0)} &= \frac{\phi_c}{M}\frac{(\rho-1)^3}{\rho}\frac{4-2\rho+\rho^2}{(4+\rho+\rho^2)^2}\,t,\\
	\hspace{-1mm}\alpha^{(1)} &= \frac{\phi_c}{K}\frac{\rho-1}{\rho(4+\rho+\rho^2)^3} \Big[6(\rho-1)^3(8+4\rho+3\rho^2)\,t \nonumber\\
   &\quad + (64+36\rho+81\rho^2+16\rho^3+18\rho^4+\rho^6)\,s\Big].
\end{align}
Consequently, $\alpha$ diverges near unjamming as $(\Delta\phi)^{-2}$. There is however a crossover from a slope $-2$ to a slope $-1.5$ at intermediate $\Delta\phi$, due to the scaling of $\nu$ with $\Delta\phi$ [Eq.~\eqref{eq:nu_Hertz}], see Fig.~\ref{bias_th}. 

For disk interacting with a harmonic potential proportional to $\delta^{2}$, Eq.~\eqref{eq:k0,mu0,k1,mu1} has coefficients $K\simeq0.3$, $s\simeq0$, $M\simeq0.2$ and $t\simeq0.5$~\cite{O'Hern03}. This translates into
\begin{equation}
	\alpha_\text{har}\sim\alpha^{(0)}(\Delta\phi)^{-3/2}+\alpha^{(1)}(\Delta\phi)^{-1},
\end{equation}
with identical $\alpha^{(0)},\alpha^{(1)}$ coefficients. At a given low $\Delta\phi$, this yields an isotropization effect that is substantial yet weaker than in the Hertzian case; see Fig.~\ref{bias_th}.

\end{document}

\widetext
\begin{center}
{\bf \Large Supporting information for ``Stress Isotropization in Weakly Jammed Granular Packings''}
\end{center}
\setcounter{section}{0}
\setcounter{equation}{0}
\setcounter{figure}{0}
\setcounter{table}{0}
\setcounter{page}{1}
\makeatletter
\renewcommand{\theequation}{S\arabic{equation}}
\renewcommand{\thefigure}{S\arabic{figure}}
\renewcommand{\theHfigure}{S\arabic{figure}}
%\renewcommand{\bibnumfmt}[1]{[S#1]}
%\renewcommand{\citenumfont}[1]{S#1}
%%%%%%%%%% Prefix a "S" to all equations, figures, tables and reset the counter %%%%%%%%%%
\vspace{5mm}

\section{Granular dynamics simulations}
We first give details on the calculation of the coarse-grained in Sec.~\ref{data_cv}. In Sec.~\ref{f_ch_plots}, we then show additional force chain snapshots of the sort presented in Fig.~3(a). In Sec.~\ref{stress_curves}, we present stress response curves as in Fig.~3(b,c) in additional parameter regimes and for isotropic forcing. Then in Sec.~\ref{finite-size}, we justify the agreement between numerical simulations and continuum theory by studying finite-size effects. Finally, in the last two sections, we slightly modify the setup. In Sec.~\ref{boundary}, we analyze the stress response at an intermediary radius between the forcing radius and the arena radius; and in Sec.~\ref{quasistatic}, we investigate the response of packings to quasistatic forcing to detect the elastic-plastic threshold.

	\subsection{Coarse-grained stresses under forcing }\label{data_cv} 
The forcing results in local stress components $\mathcal P_l\propto f_0$ and $\mathcal S_l\propto f_2$. We hereby obtain the proportionality coefficients by approximating the sum in the definition of the local stress, $\bar{\bm\sigma}^l$ [Eq.~(A2)] as an integral:
\begin{equation} 
	\bar{\sigma}^l_{ij}\approx\frac{-2D}{A_l}N_l\int_0^{2\pi} \frac{\text d\theta}{2\pi}f_i\,\hat r_j,
\end{equation}
where $\hat r_j=\delta_{jx}\cos\theta+\delta_{jy}\sin\theta$ and $N_l$ stands for the number of disks in the inner ring. Then, from the definition of the stress components [Eq.~(1)] and of the forcing [Eq.~(A1)], we obtain
\begin{subequations}
\begin{align}
	\mathcal P_l&=-\frac{\bar{\sigma}^l_{xx}+\bar{\sigma}^l_{yy}}{2}\approx \frac{DN_l}{A_l}k\int_0^{2\pi}\frac{\text d\theta}{2\pi}(f_0+2f_2\cos2\theta)(\cos^2\theta+\sin^2\theta)\approx \frac{kDN_l}{A_l} f_0,\\
	\mathcal S_l&=-\frac{\bar{\sigma}^l_{xx}-\bar{\sigma}^l_{yy}}{2}\approx \frac{DN_l}{A_l}k\int_0^{2\pi}\frac{\text d\theta}{2\pi}(f_0+2f_2\cos2\theta)(\cos^2\theta-\sin^2\theta)\approx \frac{kDN_l}{A_l} f_2.
\end{align}
\end{subequations}
Likewise, we can rewrite the stress response components from the definition of $\bar{\bm\sigma}^b$ in Eq.~(A3) by approximating the excess stress $\Delta\bm\sigma=\bm\sigma-\bm\sigma^\text{init}$ on the disks as a continuum field in polar coordinates:
\begin{subequations}
\begin{align}
	\mathcal P_b&\approx -\frac{N_b}{A_b}\frac{r_\text{out}^2}{r_\text{in}^2}\int_0^{2\pi}\frac{\text d\theta}{2\pi}\Delta\sigma_{rr}(r_\text{out},\theta),\\
	\mathcal S_b&\approx -\frac{N_b}{A_b}\frac{r_\text{out}^2}{r_\text{in}^2}\int_0^{2\pi}\frac{\text d\theta}{2\pi}\left[\Delta\sigma_{rr}(r_\text{out},\theta)\cos2\theta-\Delta\sigma_{\theta r}(r_\text{out},\theta)\sin2\theta\right],
\end{align}
\end{subequations}
where $N_b$ stands for the number of disks in the boundary ring. Therefore, by decomposing the stress response field as 
\begin{equation}
	\Delta\bm\sigma(r,\theta)=\mathbf a^{(0)}(r)+\sum_{n>0} \mathbf a^{(n)}(r)\cos n\theta+\mathbf b^{(n)}(r)\sin n\theta,
\end{equation}
it follows that only the term $a^{(0)}_{rr}$ contributes to $\mathcal P_b$, while only the terms $a^{(2)}_{rr}$ and $b^{(2)}_{\theta r}$ contribute to $\mathcal S_b$. This is in correspondence with the dependencies of the local stresses: $\mathcal P_l\propto f_0$ and $\mathcal S_l\propto f_2$.

	\subsection{Additional force chain snapshots}\label{f_ch_plots}
\begin{figure}[!h]
    \centering
    \includegraphics[width=.49\textwidth]{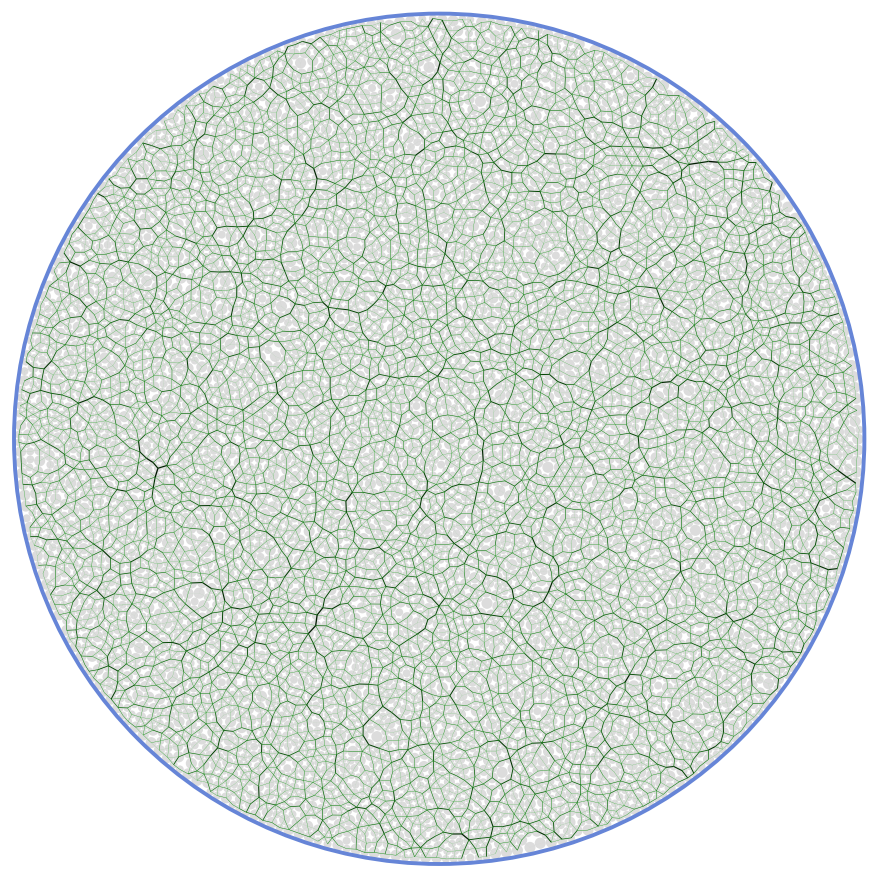}
   \caption{{\bf Isotropic force chains in an initial configuration.} Fig.~3(a) corresponds to the same realization under forcing.} 
\label{f_ch_init}
\end{figure}

\begin{figure}[!h]
    \centering
    \includegraphics[width=.49\textwidth]{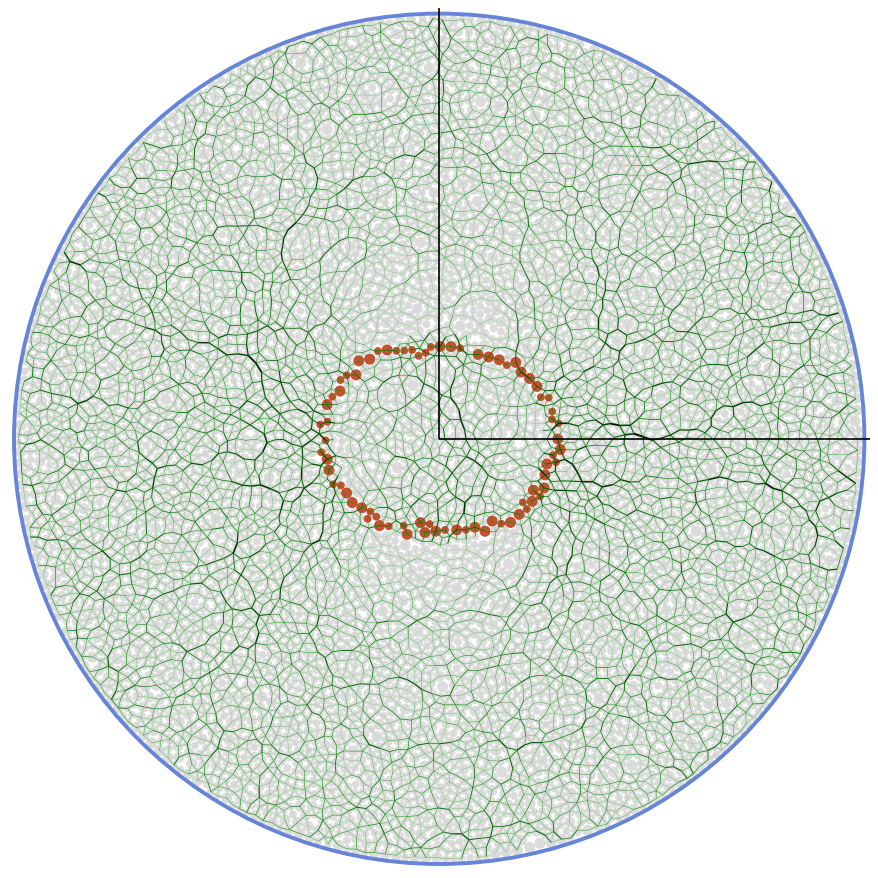}
    \hfill
    \includegraphics[width=.49\textwidth]{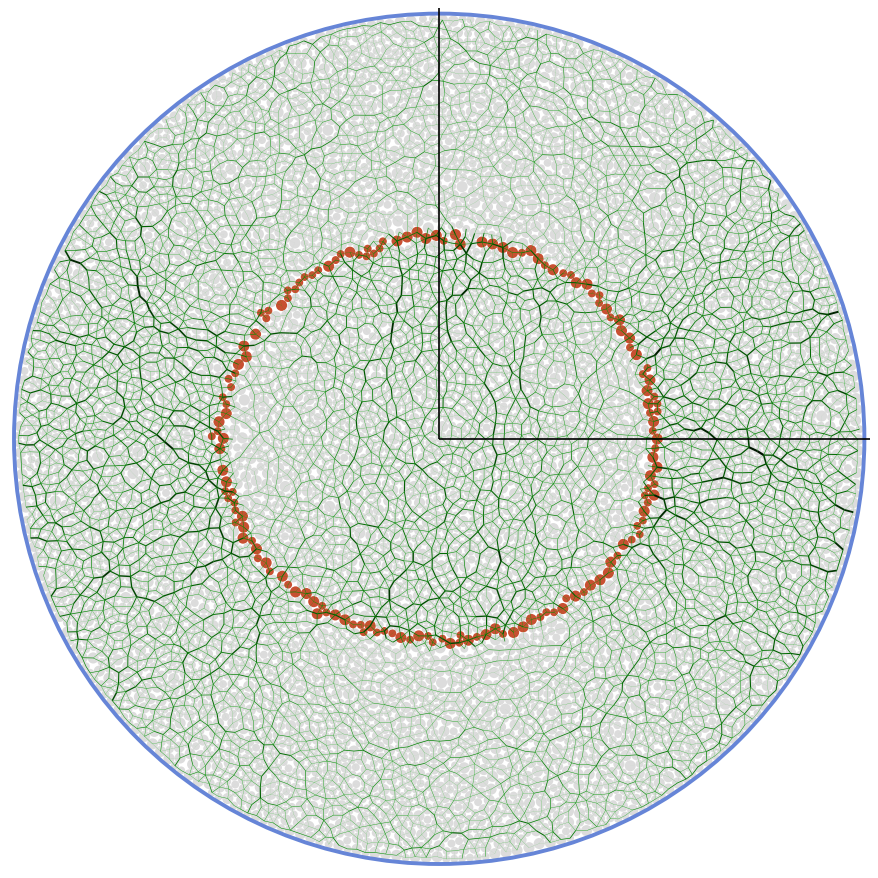}
   \caption{{\bf Force chain rearrangements for varying inner radius.} We observe similar force chain rearrangements for varying values of $r_\text{in}$. \textit{Left,} $r_\text{in}=r_\text{out}/4$. \textit{Right,} $r_\text{in}=r_\text{out}/2$. Here, $\Delta\phi \simeq 0.03$.} 
\label{f_ch_rin}
\end{figure}

\begin{figure}[!h]
    \centering
    \includegraphics[width=.49\textwidth]{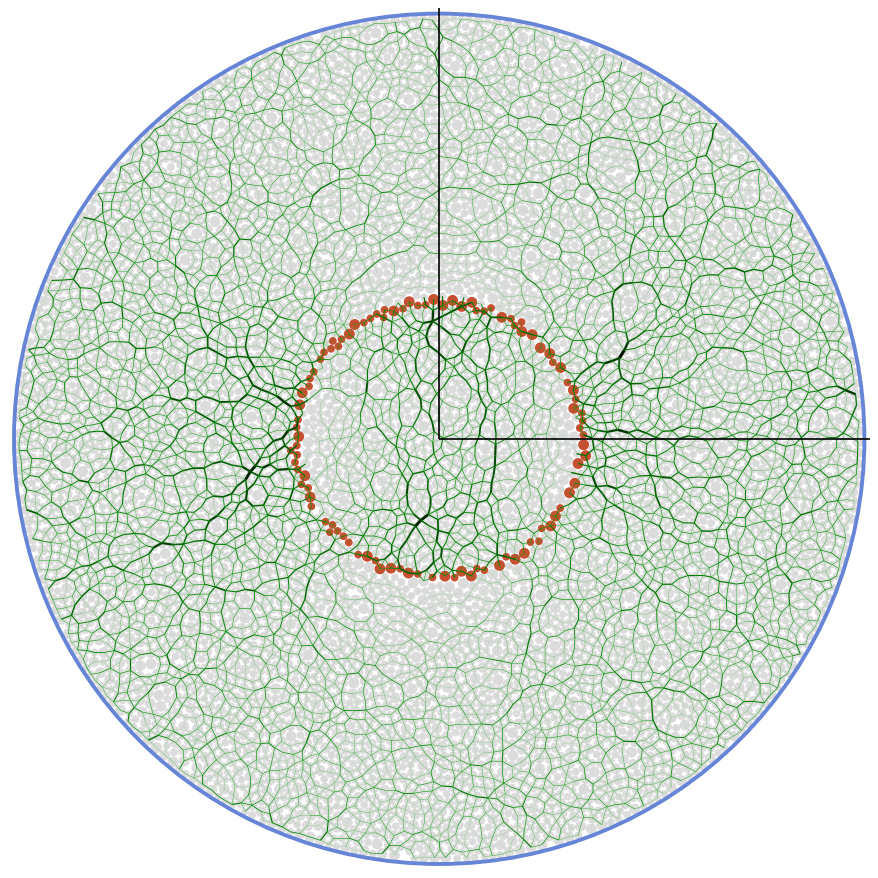}\hfill
    \includegraphics[width=.49\textwidth]{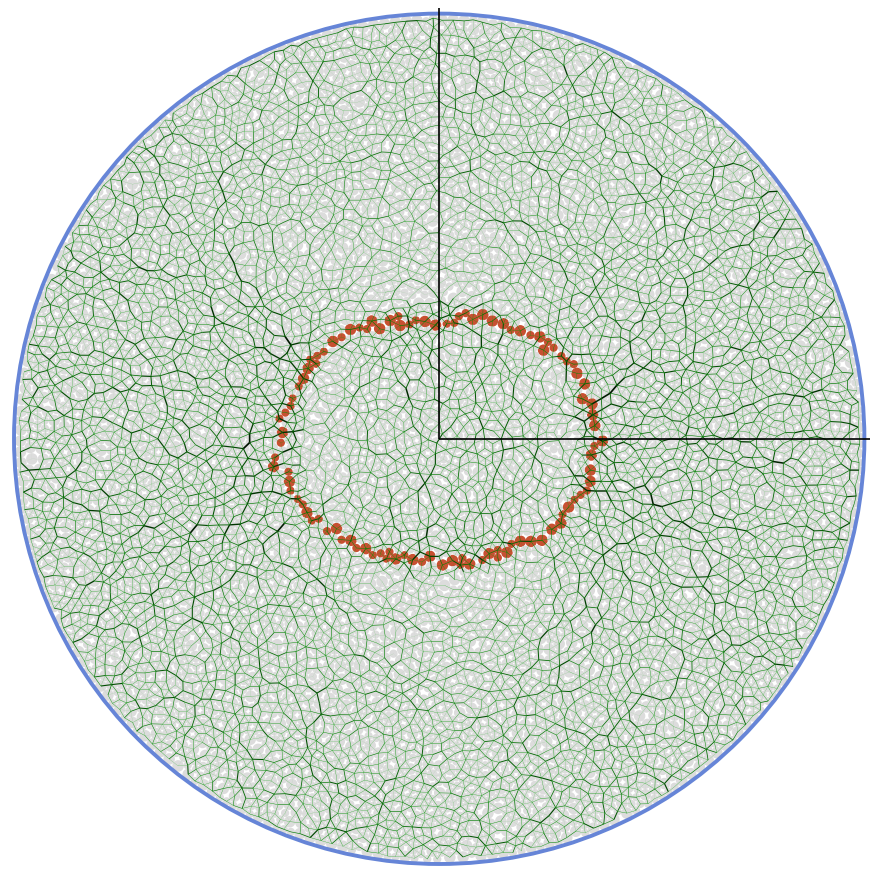}
   \caption{{\bf Force chain rearrangements for varying packing fraction.} \textit{Left,} for looser packings, we observe strengthened force chains near the boundary in all directions. Here, $\Delta\phi \simeq 0.006$ and the widths of the green segments has been artificially increased by a factor of $10$ due to the lower initial pressure, compared to Fig.~3(a). \textit{Right,} by contrast, for denser packings, we observe little force chain rearrangements. Here, $\Delta\phi \simeq 0.1$ and the widths of the green segments has been artificially decreased by a factor of $5$ due to the higher initial pressure, compared to Fig.~3(a).} 
\label{f_ch_looser_denser}
\end{figure}

Figure~\ref{f_ch_init} displays an initial configuration with isotropic force chains. Figure~\ref{f_ch_rin} shows similar force chain rearrangements as displayed in Fig.~3(a) for different values of $r_\text{in}$. When we vary the density of packings by way of $\Delta\phi=\phi-\phi_c$, we find more rearrangements for looser packings than for denser packings, see Fig.~\ref{f_ch_looser_denser}. This together validates the sketches in Fig.~1.

\newpage

$ $

\newpage

	\subsection{Additional stress response curves for moderate forcing}\label{stress_curves}
\begin{figure}[!h]
    \centering
    \includegraphics[width=.525\linewidth]{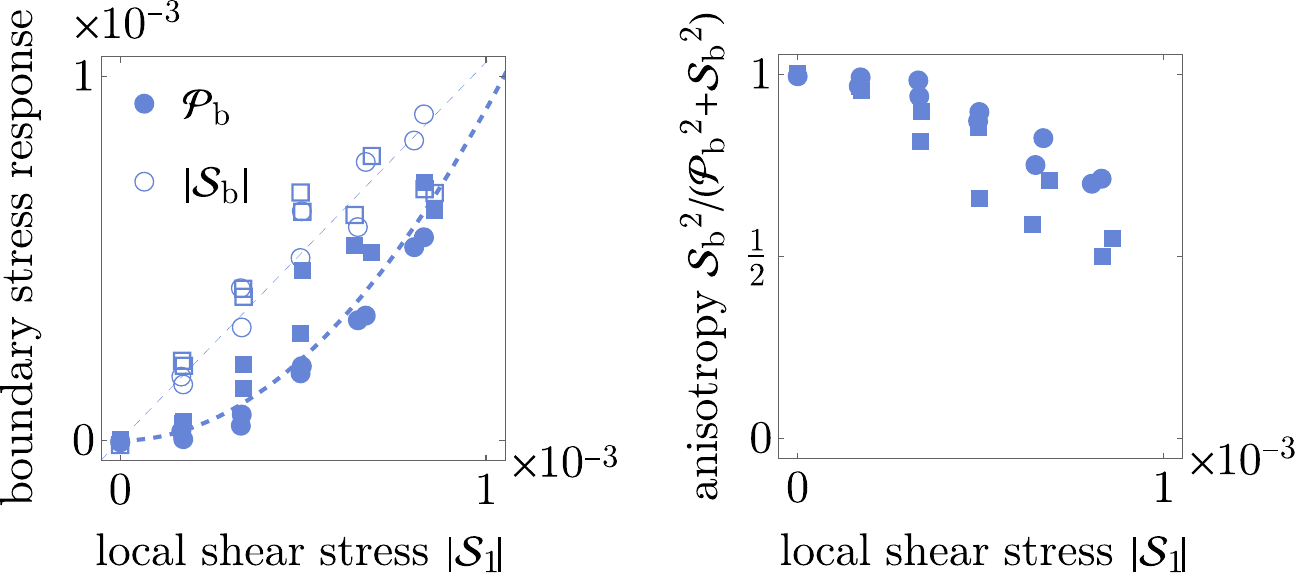}
   \caption{{\bf Weaker isotropization under dipolar forcing for $r_\text{out}/r_\text{in}=2$.} Other settings as in Fig.~3(b,c), \textit{i.e.} $r_\text{out}=44$ and $\Delta\phi\simeq 0.03$. The fits give $\alpha\simeq 910$ and $B\simeq0.04$. Here, the vertical stress response $\mathcal P_b-|\mathcal S_b|$ remains contractile (negative) for moderate forcing.}
\label{gran_dip_2}
\end{figure}

\begin{figure}[!h]
    \centering
    \includegraphics[width=.275\linewidth]{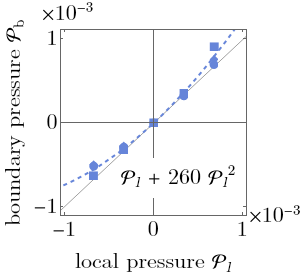}\hspace{1cm}
    \includegraphics[width=.26\linewidth]{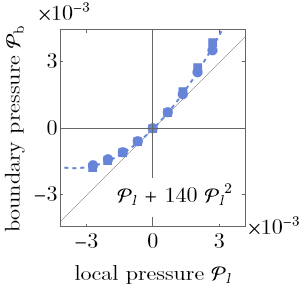}
   \caption{{\bf Dilational tendency under isotropic forcing.} The quadratic fit gives $\beta\simeq 260$ when $r_\text{out}/r_\text{in}=8$, and $\beta\simeq 140$ when $r_\text{out}/r_\text{in}=2$. Other settings as in Fig.~3(b,c).}
\label{gran_iso_8&2}
\end{figure}

\begin{figure}[!h]
    \centering
    \includegraphics[width=.56\linewidth]{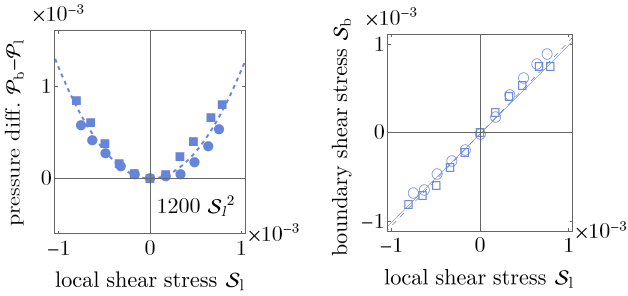}
   \caption{{\bf The dilation effects add up under dipolar plus isotropic forcing.} Same settings as in Fig.~3(b,c) except that $r_\text{out}/r_\text{in}=2$. Here, $f_0=f_2$ leads to $\mathcal P_l\approx\mathcal S_l$. The total coefficient $1200$ roughly corresponds to the sum of the coefficients under dipolar and isotropic forcing of Figs.~\ref{gran_dip_2} and \ref{gran_iso_8&2} respectively: $910+140=1150$.}
\label{gran_dipiso}
\end{figure}

Here, Fig.~\ref{gran_dip_2} displays additional results under dipolar forcing. The lower ratio of radii compared to the case of Fig.~3(b,c) yields an expectedly weaker isotropization. We then show results of simulations under isotropic forcing in Fig.~\ref{gran_iso_8&2}. Finally, Fig.~\ref{gran_dipiso} shows results under dipolar plus isotropic forcing. All data are shown for different packing histories and appear quite reproducible.

\newpage

        \subsection{Diverging length scales and finite-size effects}\label{finite-size}
Our main result consists in the quantitative agreement between the values of the isotropization coefficient $\alpha$ obtained from numerical simulations and a continuum elastic theory, as displayed in Fig.~4. Here, we review the current literature on the length scales associated with the breakdown of continuum elasticity. We then vary the total number of disks and confirm that the results of the simulations are free of finite-size effects.

%Length scales in the literature.
The first length scale we discuss here is manifested in the response to a local dipolar forcing. Due to the disordered nature of amorphous media, the displacement field around dipoles shows different spatial fluctuations in the longitudinal and perpendicular directions. These occur over length scales $\xi_L$ and $\xi_T$, respectively, both of which diverge at the unjamming transition~\citep{Silbert05,Lerner14,Karimi15}. The longitudinal length scale $\xi_L$ is not well characterized, but the transverse length scale has been obtained from simulations of harmonic spheres and disks as $\xi_T \sim(\phi-\phi_c)^{-0.24}$ for $\phi>\phi_c$~\citep{Silbert05,Lerner14}.

The second length scale we consider characterizes the response to the inflation of a single central particle. The radial fluctuations in the displacement field around the forcing occur over a length scale $l^*$ that also diverges at unjamming~\citep{Ellenbroek_PRE09}. It is the same length scale that characterizes the excess of low-frequency modes in the density of vibrational states~\citep{Wyart05,vanHecke09}. Suppose that a circular blob of radius $l$ is cut from a large rigid packing, if $l\ge l^*$, the blob should remain rigid. This leads to an estimate for $l^*$ as $1/(Z-4)\sim(\phi-\phi_c)^{-1/2}$, with $Z-4$ the excess contact number. This estimate was validated in simulations of Hertzian disks~\citep{Ellenbroek_PRE09}. The well-characterized divergence of $\xi_T$ and $l^*$ thus attest to the breakdown of continuum elasticity near unjamming.

\begin{figure}[!h]
    \centering
    \includegraphics[width=.78\linewidth]{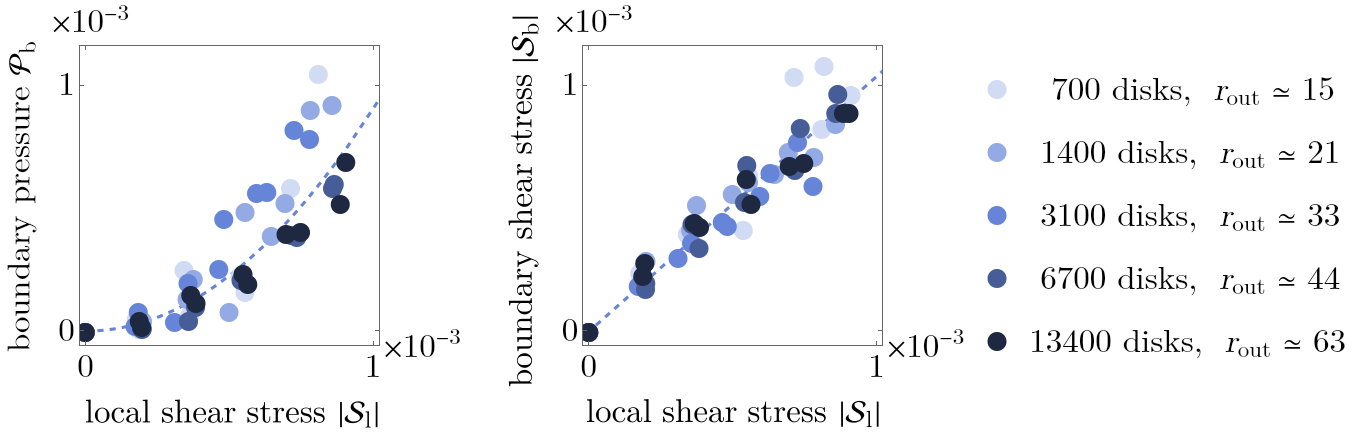}
   \caption{{\bf Robustness to system size variation.} Same settings as in Fig.~3(c,d) except that $r_\text{out}/r_\text{in}=2$ and that the total number of disks is varied. The smallest size, $700$ disks, corresponds to packings as in Fig.~2, while the largest size, $13000$ disks, corresponds to packings larger than in Fig.~3. We find no qualitative difference in the dependence of the stress response on the local shear stress. The fits are those determined in Fig.~\ref{gran_dip_2}.}
\label{gran_dip_Nb_disks}
\end{figure}

%Finite-size effects in our study. 
We now validate the notion that the systems we simulate are large enough to be consistent with the continuum elastic behavior in the limit $r_\text{out} \gg \xi_T, l^*$. To this end, we significantly vary the system size $r_\text{out}$ and assess the robustness of our prediction for the boundary stress response, $\mathcal P_b$, $\mathcal S_b$. Specifically, we vary the total number of disks from 700 to 13000, which complements our investigation of packings of 6700 disks; see the stress response curves in Fig.~\ref{gran_dip_Nb_disks}. We find no dependence of the stress response curves in Fig.~\ref{gran_dip_Nb_disks} on the value of $r_\text{out}$. This justifies the agreement between the numerical simulations and the continuum theory.\\ 

	\subsection{Stress response at an intermediary radius}\label{boundary}
The stress response could be affected by the fact that it is measured near the packing boundary. The dominant dependence on the measurement radius $r_\text{out}$ is a dilution effect, whereby stresses decrease as $r_\text{out}$ increases. We account for this stress dilution by including a factor $(r_\text{out}/r_\text{in})^2$ in the definition of $\bar{\bm\sigma}^b$, in contrast to $\bar{\bm\sigma}^l$, as detailed in Eqs.~(A2–A3). In addition, both the isotropization coefficient $\alpha$ and the phenomenological coefficient $B$ depend on the ratio $r_\text{in}/r_\text{out}$, as shown in Fig.~4(b). For fixed $r_\text{in}$, decreasing $r_\text{out}$ increases this ratio, thereby reducing $\alpha$ and hence $\mathcal{P}_b$, which enhances anisotropy in the stress response.

\begin{figure}[!h]
    \centering
    \includegraphics[width=.525\linewidth]{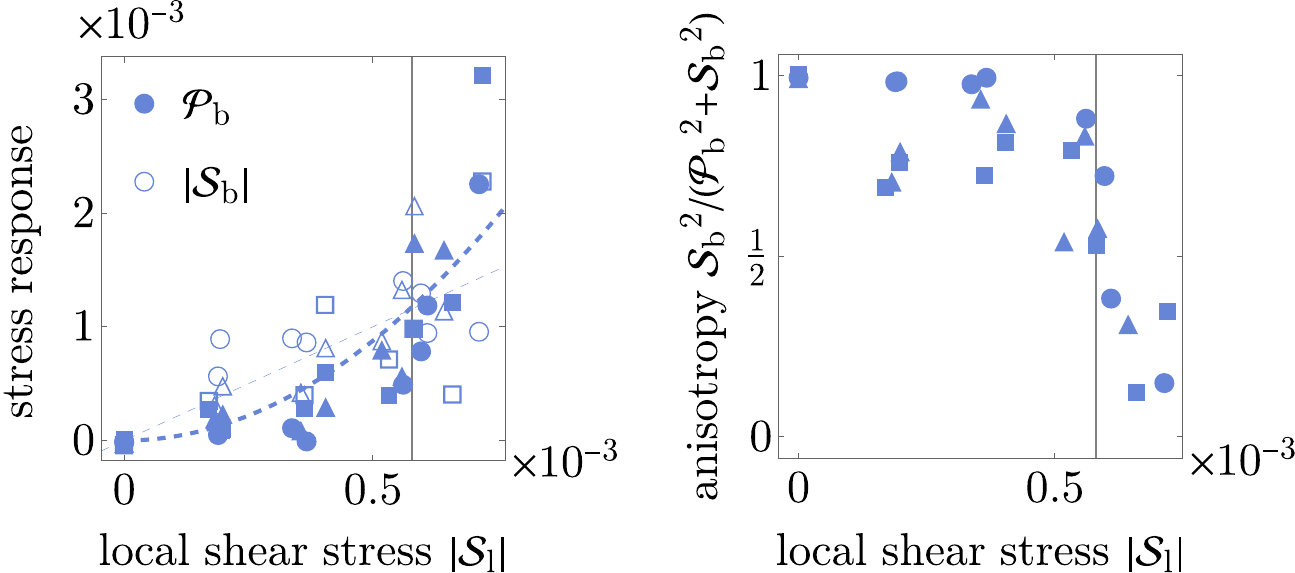}
   \caption{{\bf Robustness to variation of the measurement location.} Same settings as in Fig.~3(b,c), except that the stress response $\bar{\bm\sigma}^b$ is measured at a radius $r_\text{res}=4r_\text{in}$, inside the arena $r_\text{out}=8r_\text{in}$. As expected from Fig.~4(b), the isotropization coefficient is smaller than in Fig.~3(b): $\alpha\simeq3500$, while $B\simeq1$ is slightly larger than anticipated. Otherwise, the dependence of the stress response on the local shear stress is unchanged. The vertical line marks the local shear stress at which the anisotropy in Fig.~3(b) falls below $\frac12$.}
\label{PbvSl_rres}
\end{figure}

To further test boundary effects, we measured the stress response not at $r_\text{out}$, but at an intermediate radius $r_\text{res}$ between $r_\text{in}$ and $r_\text{out}$. As shown in Fig.~\ref{PbvSl_rres}, apart from the expected decrease in $\alpha$ and a slightly larger than expected value of $B$, the results are consistent with those in Fig.~3(b,c), obtained for $r_\text{res} = r_\text{out}$.

%\newpage

	\subsection{Quasistatic forcing and elastic-plastic threshold}\label{quasistatic}
For disk packings near unjamming, where stress isotropization is most pronounced, the yield stress vanishes~\citep{Dinkgreve15}. This raises the question of whether plastic rearrangements could drive packings significantly away from the elastic regime assumed in the derivation of the nonlinear response of Eq.~(2). To address this, we performed simulations of the packings studied in Fig.~3(b) of the main text, in which we slowly (quasistatically) increased the magnitude of the applied forcing.

\begin{figure}[!h]
    \centering
    \includegraphics[width=.85\linewidth]{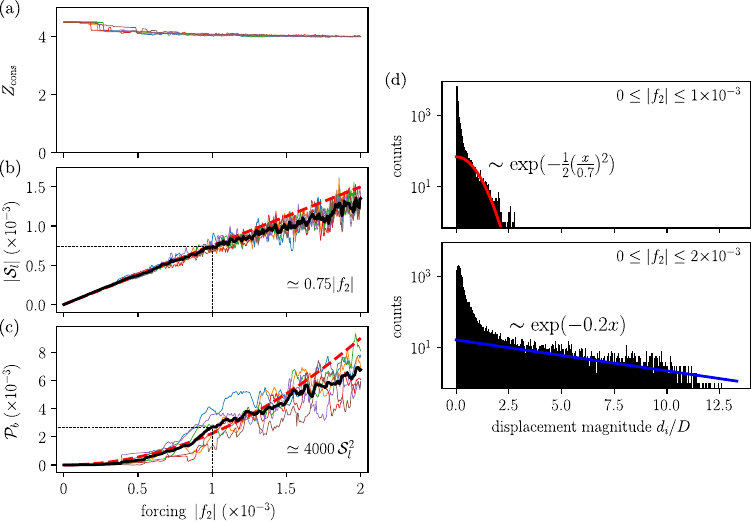}
   \caption{{\bf Response under quasistatic forcing.} 
   Same settings as in Fig.~3(b) ($r_\text{out}/r_\text{in}=8$, $r_\text{in}\simeq 5.5$, $\Delta\phi\simeq 0.03$) upon a quasistatic increase of $f_2$. 
   (a) The number of conserved contacts $Z_{\rm cons}$ decreases slightly from 4.5 to 4. 
   (b) Up to $|f_2| \simeq 1\times 10^{-3}$ (black dashed line), the local shear stress $|\mathcal S_l|$ increases linearly with only small fluctuations. Thin colored curves show the response of individual packings, the black curve their average, and the red dashed line a linear fit. 
   (c) Up to the same forcing value, the boundary pressure $\mathcal P_b$ follows the quadratic fit. 
   (d) In the elastic regime $|f_2|\le1\times10^{-3}$, disk displacements follow a Gaussian-like distribution. By contrast, the frequent avalanches for $|f_2|\ge1\times10^{-3}$ create an exponential tail. $D$ denotes the mean diameter.} 
\label{ramp_r8}
\end{figure}

If the forcing were large enough to induce significant plastic deformations (\textit{i.e.} system-spanning disk flows), we would expect major rearrangements of the contact network. To test this, we plot in Fig.~\ref{ramp_r8}(a) the average number $Z_{\rm cons}$ of contacts per disk that remain unchanged between the initial configurations and the configurations obtained after quasistatically increasing the forcing to $f_2$ (or $-f_2$). The range of $f_2$ shown extends well beyond that considered in Fig.~3(b), namely beyond $|f_2|\simeq1\times 10^{-3}$, \textit{i.e.} $|\mathcal S_l|\simeq0.75\times 10^{-3}$. We find that roughly 90\% of the initial contacts are conserved, suggesting that the contact network experiences only minor perturbations due to local rearrangements.

To check for avalanches, we next plot the local shear stress $\mathcal S_l$ and the boundary pressure $\mathcal P_b$ over the course of the same increase in $f_2$ in Figs.~\ref{ramp_r8}(b-c). The response remains fairly smooth and compatible with our nonlinear elastic prediction up to local stress values used in Fig.~3(b). For larger forcing, deviations emerge, signaling the onset of plasticity~\citep{Tanguy06,Barrat11}. To specifically test for the presence of avalanches, in Fig.~\ref{ramp_r8}(d) we plot the distribution of disk displacements between the initial configurations and the ones for $|f_2|=1\times 10^{-3}$ and $2\times 10^{-3}$. In the former case, the distribution of displacements is Gaussian-like. By contrast, in the latter case, the distribution develops exponential tails, indicative of an intermittent dynamics akin to avalanches~\citep{Maloney08}. We thus conclude that avalanches occur only under sufficiently large forcing, while the regimes explored in the main text remain relatively avalanche-free.

Finally, we note that the boundary pressure in Fig.~\ref{ramp_r8}(c) continues to qualitatively follow the elastic fit $\mathcal P_b\sim\alpha\mathcal S_l^2$ even in the avalanche-rich regime, suggesting that our conclusion remain robust beyond their nominal range of validity.

%\newpage 

\section{Mean-field elasto-plastic model} 
Here we discuss the model introduced in Eq.~(6) and the derivation of Eq.~(7) from it. We choose to base our description on the standard mean-field Hébraud-Lequeux model due to its tractability and widespread use. While this simple model does not accurately capture all quantitative aspects of the yielding transition~\cite{Nicolas18}, we merely use it to provide a proof of principle that a change in the local rules of stress propagation around a shear transformation can affect the macroscopic characteristics of this transition. Specifically, we discuss the steady state behavior of the mean stress
\begin{equation}
    \langle\sigma\rangle=\int_{-\infty}^{+\infty}\sigma P(\sigma)\,\text{d}\sigma
\end{equation}
as a function of the imposed shear rate $\dot{\gamma}$.

In the standard Hébraud-Lequeux model [the $b=0$ case of Eq.~(6)], each region of the amorphous solid drifts towards higher stresses $\sigma$ due to the externally imposed stress current $\mu\dot{\gamma}$, and starts to yield with finite probability as soon as its stress exceeds a threshold $\sigma_c$. In this section, we rescale all stresses by $\sigma_c$ and thus set $\sigma_c=1$. We denote by $\Gamma$ the rate at which shear transformations occur in the system and use the mean shear transformation time $\tau$ introduced in the main text as our time unit (\emph{i.e.}, $\tau=1$). In the Hébraud-Lequeux description, such shear transformations push the region of interest towards or away from this threshold with equal probability as implied by the $\sigma\rightarrow-\sigma$ symmetry of $a\Gamma$ diffusion term in Eq.~(6). We reason that the build-up of compressive stresses in the material resulting from stress isotropization near unjamming hampers local yielding events driven by the imposed shear stress $\mu\dot{\gamma}>0$. We thus model this effect by the simplest possible $\Gamma$-dependent drift away from the $\sigma=1$ yielding threshold, corresponding to the $b$ term of Eq.~(6).

Solving Eq.~(6) in the stationary state with boundary conditions $P(\pm\infty)=0$ as well as matching the values of $P$ and $\partial_\sigma P$ on either side of the points $\sigma=-1,0,1$ while holding $\Gamma$ constant yields a solution $P_\Gamma(\sigma)$. Imposing the normalization condition $\int_{-\infty}^{+\infty}P_\Gamma(\sigma)\,\text{d}\sigma=1$ on this solution yields the following self-consistency condition on $\Gamma$: 
\begin{equation}\label{eq:self-consistent}
    a =  \frac{x^2}{2}+y^{-1}\left[1+x\sqrt{1+\left(\frac{xy}{2}\right)^2}\right]
    \frac{\frac{xy}{2}+\sqrt{1+\left(\frac{xy}{2}\right)^2}\tanh{\frac{y}{2}}}{\frac{xy}{2}\tanh\frac{y}{2}+\sqrt{1+\left(\frac{xy}{2}\right)^2}},
\end{equation}
where we have defined
\begin{equation}
    x = \sqrt{a\Gamma}
    \qq{and}
    y = \frac{\mu\dot{\gamma}}{a\Gamma}-b.
\end{equation}
For a given externally imposed shear rate $\dot{\gamma}$, solving Eq.~\eqref{eq:self-consistent} yields a value for $\Gamma$, and thus information about the flow state of the system.

\begin{figure}
    \centering
    \includegraphics[width=\linewidth]{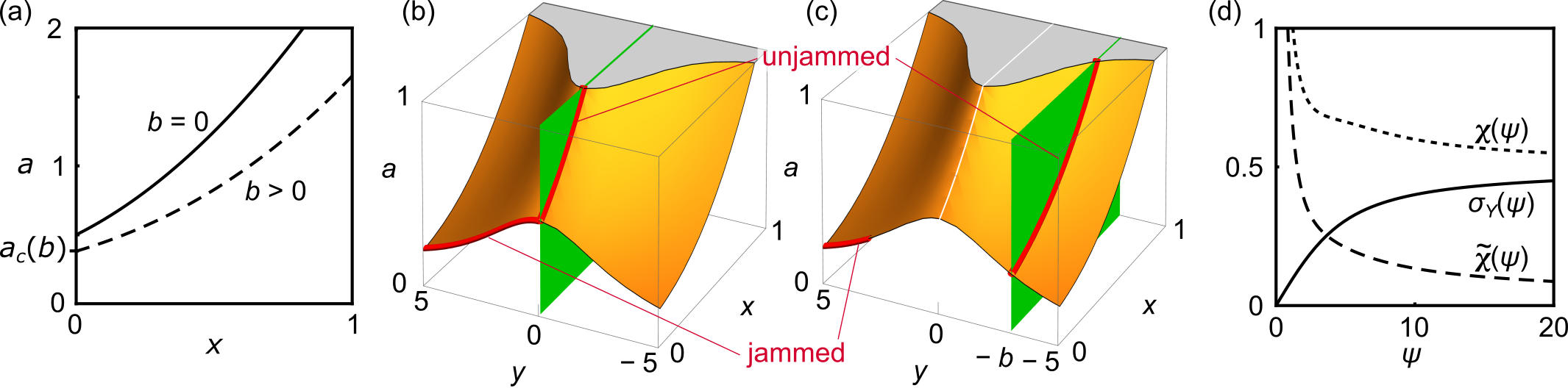}
    \caption{
    {\bf Graphical illustration of the nonlinear equations involved in the mean-field elastoplastic model.}
    (a)~No-flow, liquid-state self-consistency condition obtained by setting $\mu\dot\gamma=0$ (\emph{i.e.}, $y=-b$) in Eq.~\eqref{eq:self-consistent}. 
    For a given value of the parameter $a$, this curve allows us to read out the corresponding value of $x$.
    (b)~Illustration of the change of solution of Eq.~\eqref{eq:self-consistent} when transitioning from an unjammed (liquid-like) to a jammed (solid-like) state in the standard Hébraud-Lequeux $b=0$ case. The yellow surface materializes the solutions of Eq.~\eqref{eq:self-consistent}. For a given value of $a$, all the solutions in the plane with altitude $a$ are accessible. Those with zero flow are given by the intersection from this yellow plane with the green $\dot\gamma=0\Leftrightarrow y=0$ plane (``unjammed'' red line). Imposing a flow in the $\dot\gamma>0$ direction makes this solution move away from the green plane and along the line defined by the intersection of the yellow surface and the horizontal plane (\emph{i.e.}, the plane of constant $a$). In cases where $a<1/2$, a change of sign of $\dot\gamma$ is accompanied by a discontinuous jump of the value of $y$. The $\dot\gamma=0^+$ solutions are marked by the ``jammed'' red line.
    (c)~Selection of the solutions in the $b>0$ case. Starting from a large value of $a$, we can follow the ``unjammed'' red line at the intersection of the yellow surface and the $y=-b$ plane. This line ends at $a=a_c(b)$, which marks the jamming transition. This value of $a$ is smaller than in the $b=0$ case, and does not correspond to the top of the bump in the yellow surface. For values of $a$ lower than $a_c(b)$, the $\dot\gamma>0$ branch is obtained by jumping to the $y>0$ part of the yellow surface. Unlike in the $b=0$ case, such a jump is required even in the critical case $a=a_c(b)$. The line of jammed solutions highlighted in red gives the $\dot\gamma\rightarrow 0^+$ limit of these solutions. The discussion of the text is focused on the neighborhood of this line.
    (d)~Plots of the quantities defined in Eqs.~\eqref{eq:yield_stress_susceptibility} and \eqref{eq:critical_rheology}.
    }
    \label{fig:HL_curves}
\end{figure}

We first consider the case of a system without externally imposed flow ($\dot{\gamma}=0$). The value of $y$ is then known and Eq.~\eqref{eq:self-consistent} must be solved for $x$. As shown if Fig.~\ref{fig:HL_curves}(a), it allows for $x>0$ solutions only when $a$ is larger than a critical value 
\begin{equation}
    a_c(b)=\frac{1}{b}\tanh\frac{b}{2}.
\end{equation}
The parameter $a$ was originally introduced in Ref.~\cite{Hebraud98} as a proxy for the density of the system, where denser systems correspond to lower values of $a$. This gives rise to the following standard interpretation for the existence of a $x>0$ solution to Eq.~\eqref{eq:self-consistent}:
\begin{itemize}
    \item Parameter regimes where this solution exists correspond to low-density systems, namely those with $a>a_c(b)$. Such systems display a non-vanishing steady-state value for their plastic activity $\Gamma$. In the case $b=0$, such systems are usually interpreted as unjammed fluids, and we extend this interpretation to the case $b>0$.
    \item Parameter regimes where this solution does not exist represent high-density systems, namely those with $a<a_c(b)$. These systems display a vanishing value of $\Gamma$ in the case of a vanishing drive $\mu\dot{\gamma}=0$. Imposing a small, positive value of $\mu\dot{\gamma}$ does however result in a non-vanishing $\Gamma$ and a discontinuous jump in the mean stress $\langle\sigma\rangle$, likening systems in this regime to jammed yield-stress fluids.
\end{itemize}

In our model, the former of these two cases gives rise to a linear, Newton-like rheology similar to that derived from the standard Hébraud-Lequeux model. However, as the parameter $a$ is lowered under $a_c(b)$, this solution ceases to exist and the system transitions to another solution of Eq.~\eqref{eq:self-consistent} [Fig.~\ref{fig:HL_curves}(b-c)]. For finite values of $b$, the $\dot\gamma\rightarrow 0^+$ asymptotics of this solution is given by
\begin{subequations}
    \begin{align}
        x &= \sqrt{\frac{\mu\dot{\gamma}}{b+\psi}} + \mathcal{O}(\mu\dot{\gamma})\\
        y &= \psi\left(1+\frac{\sinh\psi+\psi}{\sinh\psi-\psi}\times\sqrt{\frac{\mu\dot{\gamma}}{b+\psi}}\right) + \mathcal{O}(\mu\dot{\gamma}),
    \end{align}
\end{subequations}
where the function $\psi(a)$ depends only on parameter $a$ and is defined through
\begin{equation}
    a = \frac{1}{\psi}\tanh\frac{\psi}{2} \qquad \text{and} \qquad \psi>0.
\end{equation}
Thus, $\psi$ is equal to $0$ when $a=1/2$ and monotonically increases to $+\infty$ as $a$ decreases to $0$. The resulting relationship between stress and strain rate is of the yield stress type, with
\begin{equation}
    \langle\sigma\rangle=\sigma_Y(\psi) + \frac{\chi(\psi)}{\sqrt{b+\psi}} \sqrt{\mu\dot\gamma} + \mathcal{O}(\mu\dot{\gamma}),
\end{equation}
where the yield stress and susceptibility are respectively given by
\begin{subequations}\label{eq:yield_stress_susceptibility}
    \begin{align}
        \sigma_Y(\psi) &= \frac{1}{2}\coth\frac{\psi}{2}-\frac{1}{\psi}\\
        \chi(\psi) &= \frac{1}{\psi} + \frac{5+\cosh\psi-3\psi\coth(\psi/2)}{2(\sinh\psi-\psi)}
    \end{align}
\end{subequations}
and are plotted in Fig.~\ref{fig:HL_curves}(d).

As illustrated in Fig.~\ref{fig:HL_curves}(b-c), unlike in the Hébraud-Lequeux model, the transition between the unjammed and jammed state in our $b>0$ model is of the first order. This is in contrast with the transition encountered in the Hébraud-Lequeux model, which is of the second order and displays a critical regime where $\langle\sigma\rangle\propto(\dot\gamma)^{1/5}$. This critical regime is never manifested in our $b>0$ model, and the rheology at the transition is given by (noting that $a=a_c\Rightarrow\psi=b$):
\begin{equation}\label{eq:critical_rheology}
    \langle\sigma\rangle_\text{at unjamming}=\sigma_Y(b) + \tilde{\chi}(b)\sqrt{\mu\dot\gamma} + \mathcal{O}(\mu\dot{\gamma}),
\end{equation}
where $\tilde{\chi}(b)=\chi(b)/\sqrt{2b}$ is plotted in Fig.~\ref{fig:HL_curves}(d).
According to Eq.~\eqref{eq:critical_rheology}, at the unjamming transition the $b>0$ system displays the same yielding behavior and Herschel-Bulkley exponent as in the jammed state. 
The small-$\dot\gamma$ expansion scheme leading to Eq.~\eqref{eq:critical_rheology} breaks down at $b=0$, which is apparent in the divergence of $\tilde{\chi}(b)$ for $b\rightarrow 0^+$. Consistent with our previous discussion, this divergence signals the presence of the Hébraud-Lequeux critical regime. Equation~(7) of the main text is a summarized version of Eq.\eqref{eq:critical_rheology}.

We note that the difference in critical regime discussed here is fundamentally due to a loss of the $\sigma\rightarrow -\sigma$ symmetry in our model, which implies a change in the order of the unjamming transition. We can speculate that some version of this difference will subsist even in more sophisticated models, \emph{i.e.}, that it is a robust feature beyond the assumptions and simplifications made here. We however expect that the specific values of the Hershel-Buckley exponents discussed here will change when, \emph{e.g.}, relaxing our mean-field assumption. 

%\bibliography{bibfile0}

%